\documentstyle{elsart}

\def\bra#1{\mbox{$\langle #1\vert $}}
\def\ket#1{\mbox{$\vert #1\rangle$}}

\begin{document}
\begin{frontmatter}

%\centerline{\Large{\bf Nucleon Magnetic Moments in an }}
%\medskip
%\centerline{\Large{\bf Extended Chiral Constituent Quark Model}}
\title{Nucleon Magnetic Moments in an \\ 
Extended Chiral Constituent Quark Model}

%\vskip 1.5cm

%\centerline{\large{R.F.~Wagenbrunn$^{M.~Radici$^{1,2}$,
%S.~Boffi$^{1,2}$ and P.~Demetriou$^{3}$}}
\author[1]{R.F.~Wagenbrunn}, 
\author[1,2]{M.~Radici}, 
\author[1,2]{S.~Boffi} and 
\author[3]{P.~Demetriou}

%\vskip 1.0cm

%\centerline{\small $^1$~Dipartimento di Fisica Nucleare e Teorica,
%Universit\`a di Pavia, and}
%\centerline{\small $^2$~Istituto Nazionale di Fisica Nucleare, 
%Sezione di Pavia, Pavia, Italy}
%\centerline{\small $^3$~Institute of Nuclear Physics, N.C.S.R.
%``Demokritos'', Athens, Greece}
\address[1]{Dipartimento di Fisica Nucleare e Teorica,
Universit\`a di Pavia, Italy}
\address[2]{Istituto Nazionale di Fisica Nucleare, 
Sezione di Pavia, Pavia, Italy}
\address[3]{Institute of Nuclear Physics, N.C.S.R.
``Demokritos'', Athens, Greece}

%\vskip 1.5cm

\begin{abstract}

We present results for the nucleon magnetic moments in the context of an
extended chiral constituent quark model based on the mechanism of the
Goldstone boson exchange, as suggested by the spontaneous breaking of
chiral symmetry in QCD. The electromagnetic charge-current
operator is consistently deduced from the model Hamiltonian, which
includes all force components for the pseudoscalar, vector and scalar meson
exchanges. Thus, the continuity equation is satisfied for each piece of the
interaction, avoiding the introduction of any further parameter. A
good agreement with experimental values is found. The role of isoscalar
two-body operators, not constrained by the continuity equation, is also
investigated.

\vspace{.2cm}

\end{abstract}

%{\sl Keywords\/}: constituent quark models, gauge-invariant e.m. currents,
%nucleon magnetic moments.
%\begin{keyword}
%
%Constituent quark models. Gauge-invariant e.m. currents. Nucleon magnetic 
%moments.
%
%\end{keyword}

\end{frontmatter}

In the context of a Constituent Quark Model (CQM) based on the Goldstone
Boson Exchange (GBE) mechanism (GBE CQM), the spectrum of light and 
strange baryons can be successfully reproduced in a unified manner retaining 
just the spin-spin part of the interaction represented by the exchange of 
the pseudoscalar meson octet~\cite{GBECQM,GBECQM1}. However, the
electromagnetic two-body operator, that can be consistently deduced by
satisfying the continuity equation, has been shown to give no contribution
to, e.g., the nucleon magnetic moments, thus leading to an underestimation of
their measured values~\cite{prc}.

In the socalled extended GBE CQM, the missing tensor force of the
pseudoscalar $(\pi,K,\eta,\eta')$ exchange has been introduced, as well as 
multiple GBE through the exchange of vector $(\rho,K^*,\omega_8,\omega_0)$
and scalar $(\sigma)$ mesons. A baryon spectrum of quality
comparable to the GBE CQM has been obtained~\cite{exGBECQM,exGBECQM1}. The 
Hamiltonian is given by 
\begin{eqnarray}
H &= &\sqrt{{\cal M}^2 + \vec P^{\,2}} \nonumber \\
{\cal M} &= &\sum_{i=1}^3 \sqrt{{\vec y}^{\,2}_i + m_i^{2}} + \sum_{i<j=1}^3 
V_{ij} \, ,
\label{eq:hamiltonian}
\end{eqnarray}
where $\vec P=\displaystyle{\sum_{i=1}^3} \, \vec p_i$ is the center-of-mass 
(cm) momentum of the three constituent quarks with mass $m_i$ and momentum 
$\vec p_i$, while the mass operator ${\cal M}$ describes the intrinsic motion 
of the quarks inside the baryon in terms of their intrinsic momenta $\vec y_i = 
\vec p_i -\textstyle{1\over3} \vec P$ and mutual interaction $V_{ij}$. 
The use of the relativistic expression 
for the kinetic energy operator avoids the typical drawback of the 
nonrelativistic CQM, where the mean velocity of quarks can become larger than 
the velocity of light. 

As a further test of the extended GBE CQM, we will consider here the magnetic 
moments of the nucleon by calculating the matrix elements of the electromagnetic 
charge-current operator deduced consistently from the Hamiltonian $H$ of Eq.
(\ref{eq:hamiltonian}). In this way, the charge-current operator is gauge 
invariant, satisfies the continuity equation and no further parameters are 
introduced with respect to the extended GBE CQM. The initial and final states 
$(\ket{i}, \ket{f})$ are taken as the factorized product of the eigenfunctions
of ${\cal M}$ and of plane waves for the cm motion.

The electromagnetic charge-current operator consists of a one- and a
two-body part. Since the kinetic energy operator of the cm and intrinsic motion 
in Eq. (\ref{eq:hamiltonian}) contains the square root operator and the potential
is local, a
gauge invariant one-body operator is deduced by applying the minimal 
substitution with an external electromagnetic field $A=(A_0, \vec A)$ and 
then by using the formalism of the functional derivation. In fact, using the 
results of Ref.~\cite{npa}, we can define
\begin{eqnarray}
H(A) &= &\sqrt{{\cal M}^2(\vec A) + \left[ \sum_i \left( \vec p_i - e_i 
\vec A \right) \right]^2} + e_N A_0 \nonumber \\
&\equiv &\sqrt{{\cal M}^2(\vec A) + R^2(\vec A)} + e_N A_0 \, ,
\label{eq:minsubst}
\end{eqnarray}
where $e_i$ are the individual quark charges and $e_N$ is the nucleon
charge (both expressed in units of the proton charge), and then represent 
the matrix element of the one-body charge-current operator as
\begin{eqnarray}
J_0 &= 
&\int d \vec x \e^{\mathrm{i}\vec q \cdot \vec x} 
\bra{f} {{\delta H(A)} \over {\delta A_0(\vec x)}} \Bigg \vert_{A_0=0} \ket{i} 
= e_N \, \delta (\vec P' - \vec P -\vec q) \label{eq:charge} \\
\vec J^{\mathrm{drift}} &= 
&\int d \vec x \e^{\mathrm{i}\vec q \cdot \vec x} 
\bra{f} {{\delta H(A)} \over {\delta \vec A(\vec x)}} \Bigg \vert_{\vec A=0} 
\ket{i} = 
{1\over{E+E'}} \int d \vec x \e^{\mathrm{i}\vec q \cdot \vec x} 
\bra{f}{{\delta H^2(A)} \over {\delta \vec A(\vec x)}} \Bigg \vert_{\vec A=0} 
\ket{i} \nonumber \\
&= &{{2M}\over{E+E'}} \int d \vec x \e^{\mathrm{i}\vec q \cdot \vec x} 
\bra{f}{{\delta {\cal M}(A)} \over {\delta \vec A(\vec x)}}
\Bigg \vert_{\vec A=0} \ket{i} 
+ e_N \, {{\vec P + \vec P'}\over{E+E'}} \, \delta(\vec P' - \vec P - \vec q) 
\nonumber \\
&\equiv &{{2M}\over{E+E'}} \vec J_{\mathrm{intr}}^{\mathrm{drift}}  
+ e_N \, {{\vec P + \vec P'}\over{E+E'}} \, \delta (\vec P' - \vec P - \vec q) 
\, . \label{eq:drift}
\end{eqnarray}
Here, $\vec q$ is the momentum transferred by the external field at the
space point $\vec x$, $M$ is the nucleon mass and $\vec P (\vec P'), E 
(E')$ are the cm momentum and total energy of the initial (final) state 
$\ket{i(f)}$, respectively. The spatial part (\ref{eq:drift}) 
represents the contribution of the total drift current: it contains a cm 
part, that describes the nucleon as a whole, and a part 
$\vec J_{\mathrm{intr}}^{\mathrm{drift}}$ related to the intrinsic motion. 
The latter can be made explicit by again systematically applying the minimal 
substitution to each quark momentum variable and then using the techniques 
of functional derivation~\cite{npa}:
\begin{eqnarray}
\vec J_{\mathrm{intr}}^{\mathrm{drift}} &= 
&\int d \vec x \e^{\mathrm{i}\vec q \cdot \vec x} 
\bra{f} \sum_{l=1}^3 {\delta \over {\delta \vec A_l(\vec x)}} \nonumber \\
&{}&\! \! \! \! \! \! \! \! \! \! \! \! \! \! \! \! \! \! \! \! \! \! \! \!
\sum_{i=1; \, j,k\ne i}^3 
\sqrt{ \left[ {2\over3} \left( \vec p_i - e_i \vec A_i (\vec x) \right) 
- {1\over3} \left( \vec p_j - e_j \vec A_j (\vec x) \right) - 
{1\over3} \left( \vec p_k - e_k \vec A_k (\vec x) \right) \right]^2 
+ m_i^2} \Bigg \vert_{\vec A_l=0} \ket{i} \nonumber \\
&= &\bra{f} \sum_{i=1; \, j,k\ne i}^3 {{\vec y_i + \vec y_i^{\, \prime}}
\over{E_i+E_i'}} 
\, \left[ {2\over3} e_i \, \delta (\vec p_i^{\, \prime} -\vec p_i -\vec q) \, 
\delta (\vec p_j^{\, \prime} -\vec p_j) \, \delta (\vec p_k^{\, \prime} -
\vec p_k) \right. \nonumber \\
&{}&\qquad \qquad \qquad \qquad \left. \begin{array}{l}  -
\displaystyle{{1\over3}} e_j \, 
\delta (\vec p_j^{\, \prime} -\vec p_j -\vec q) \, 
\delta (\vec p_k^{\, \prime} -\vec p_k) \, 
\delta (\vec p_i^{\, \prime} -\vec p_i) \end{array} \right. \nonumber \\
&{}&\qquad \qquad \qquad \qquad \left. \begin{array}{l} -
\displaystyle{{1\over3}} e_k \, 
\delta (\vec p_k^{\, \prime} -\vec p_k -\vec q) \, 
\delta (\vec p_i^{\, \prime} -\vec p_i) \, 
\delta (\vec p_j^{\, \prime} -\vec p_j) \end{array} \right] \ket{i} \, ,
\label{eq:driftintr}
\end{eqnarray}
where $E_i=\sqrt{\vec y_i^{\,2} +m_i^2}$ and $\vec y_i^{\, \prime}=
\vec y_i+ \textstyle{{2\over3}} \vec q$. 

Following the lines of Ref.~\cite{npa}, the one-body spin magnetic current can 
also be deduced by applying the minimal substitution to the equivalent 
Hamiltonian 
\begin{eqnarray}
H &= &\sqrt{{\cal M}_s^2 + (\vec \sigma \cdot \vec P)^2} \equiv 
\sqrt{{\cal M}_s^2 + R_s^2} \nonumber \\
{\cal M}_s &= &\sum_{i=1}^3 \sqrt{(\vec \sigma \cdot \vec y_i)^2 + m_i^2} + 
\sum_{i<j=1}^3 V_{ij} 
\label{eq:spinhamilt}
\end{eqnarray}
and by defining 
\begin{eqnarray}
\vec J^{\mathrm{spin}} &= &\vec J^{\mathrm{tot}} - \vec J^{\mathrm{drift}}
\nonumber \\
&= &\int d \vec x \e^{\mathrm{i}\vec q \cdot \vec x} 
\bra{f} {\delta \over {\delta \vec A(\vec x)}} \left[ 
\sqrt{{\cal M}_s^2(A) + R_s^2(A)} - \sqrt{{\cal M}^2(A) + R^2(A)} \right] 
\Bigg \vert_{\vec A=0} \ket{i} \nonumber \\
&= &\int d \vec x \e^{\mathrm{i}\vec q \cdot \vec x} 
\left\{ {1\over{E+E'}} \bra{f} {\delta \over {\delta \vec A(\vec x)}} \left[
R_s^2(A) - R^2(A) \right] \Bigg \vert_{\vec A=0} \ket{i} \right. \nonumber \\
&{}&\left. \qquad \qquad + {2M\over{E+E'}} \bra{f} {\delta \over {\delta 
\vec A(\vec x)}} \left[ {\cal M}_s(A) - {\cal M}(A) \right] \Bigg 
\vert_{\vec A=0} \ket{i} \right\} \nonumber \\
&\equiv &\vec J^{\mathrm{spin}}_{\mathrm{cm}} + 
\vec J^{\mathrm{spin}}_{\mathrm{intr}} \, .
\label{eq:spin}
\end{eqnarray}
After some algebra, the final result for the cm and intrinsic one-body spin
currents is
\begin{eqnarray}
\vec J^{\mathrm{spin}}_{\mathrm{cm}} &= &
{{\mathrm{i}}\over{E+E'}} \bra{f} 
\sum_{i=1; \, j,k\ne i}^3 \, e_i \, \vec \sigma_i \times 
(\vec p_i^{\, \prime} - \vec p_i) \nonumber \\
&{}&\qquad \qquad \qquad \qquad 
\delta (\vec p_i^{\, \prime} -\vec p_i -\vec q) \, 
\delta (\vec p_j^{\, \prime} -\vec p_j) \, 
\delta (\vec p_k^{\, \prime} -\vec p_k) \ket{i} \label{eq:spincm} \\
\vec J^{\mathrm{spin}}_{\mathrm{intr}} &= &
{\mathrm{i}} \, \bra{f} 
\sum_{i=1; \, j,k\ne i}^3 {1\over{E_i+E_i'}} \nonumber \\
&{}&\quad \left[ {4\over9} e_i \, \vec \sigma_i \times 
(\vec p_i^{\, \prime} - \vec p_i) \, 
\delta (\vec p_i^{\, \prime} -\vec p_i -\vec q) \, 
\delta (\vec p_j^{\, \prime} - \vec p_j) \, 
\delta (\vec p_k^{\, \prime} -\vec p_k) \right. \nonumber \\
&{}&\left. \quad \begin{array}{l} + 
\displaystyle{{1\over9}} e_j \, \vec \sigma_j \times 
(\vec p_j^{\, \prime} - \vec p_j) \, 
\delta (\vec p_j^{\, \prime} -\vec p_j -\vec q) \, 
\delta (\vec p_k^{\, \prime} -\vec p_k) \, 
\delta (\vec p_i^{\, \prime} -\vec p_i) \end{array} \right. 
\nonumber \\
&{}&\left. \quad \begin{array}{l} + 
\displaystyle{{1\over9}} e_k \, \vec \sigma_k \times 
(\vec p_k^{\, \prime} - \vec p_k) \, 
\delta (\vec p_k^{\, \prime} -\vec p_k -\vec q) \, 
\delta (\vec p_i^{\, \prime} -\vec p_i) \, 
\delta (\vec p_j^{\, \prime} -\vec p_j) \end{array} \right]
\ket{i} \, , \label{eq:spinintr}
\end{eqnarray}
where $\vec \sigma_i$ means that the matrix element is taken on the spin of
the $i$-th quark. 

The two-body part of the electromagnetic current operator can be derived
directly from the continuity equation 
\begin{eqnarray}
\vec q \cdot \left( \vec J_{\left[ 1 \right]}^{\mathrm{cm}} + 
\vec J_{\left[ 1 \right]}^{\mathrm{intr}} + 
\vec J_{\left[ 2 \right]} \right) &= &\bra{f} \left[ H, J^0_{\left[ 1 \right]} 
\right] \ket{i} = {1\over{E+E'}} \bra{f} \left[ {\cal M}_s^2 + R_s^2, 
J^0_{\left[ 1 \right]} \right] \ket{i} \nonumber \\
&{}&\! \! \! \! \! \! \! \! \! \! \! \! \! \! \! \! \! \! \! \! \! \! \! \!
\! \! \! \! \! \! \! \! \! \! \! \! \! \! \! \! \! \! \! \! \! \! \! \!
\! \! \! \! \! \! \! \! \! \! \! \! \!\! \! \! \! \! \! \! 
\begin{array}{l} = \displaystyle{{1\over{E+E'}}} \bra{f} 
\left[ R_s^2, J^0_{\left[ 1 \right]} \right] \ket{i} + 
\displaystyle{{2M\over{E+E'}}} \left( \bra{f} 
\left[ {\cal M}_s - V , J^0_{\left[ 1 \right]} \right]  + 
\left[ V , J^0_{\left[ 1 \right]} \right] \ket{i} \right) \end{array}
\label{eq:cont}
\end{eqnarray}
consistently with the Fourier transform of the potential in Eq. 
(\ref{eq:hamiltonian}) and of the one-body charge operator. Here, we will 
consider only the SU(2) sector of chiral symmetry, neglecting the strange 
quark. Therefore, the flavor (isospin) dependence of the charge generates 
non-vanishing exchange currents related to $\pi$ and $\rho$ exchanges only. In 
particular, the pseudoscalar piece gives the well known isovector 
pion-pair $(\pi q \overline{q})$ and pion-in-flight $(\gamma \pi \pi)$ 
currents~\cite{ts99}
\begin{eqnarray}
\vec J_{\pi q\bar q}(\vec k_i, \vec k_j) &= &{\mathrm{i}} \, {{g_{\pi}^2}\over 
{4m_i m_j}} \left[ 
\frac{{\vec\sigma}_i\cdot {\vec k}_i}{({\vec k}_i^2+m_{\pi}^2)}\,{\vec\sigma}_j
\left(\frac{\Lambda_{\pi}^2-m_{\pi}^2}{{\vec k}_i^2+\Lambda_{\pi}^2}\right)^2\,
-\, ( i \leftrightarrow j )\right] \nonumber \\
&{}&\qquad ({\vec \tau}_i \times {\vec \tau}_j)_z 
\label{eq:piqq}
\end{eqnarray}
\begin{eqnarray}
\vec J_{\gamma \pi\pi}(\vec k_i, \vec k_j) &=& {\mathrm{i}} \, 
\frac{g_{\pi}^2}{4m_i m_j}
\frac{{\vec\sigma}_i\cdot {\vec k}_i\,{\vec\sigma}_j\cdot {\vec k}_j}
{({\vec k}_i^2+m_{\pi}^2)({\vec k}_j^2+m_{\pi}^2)}\,({\vec k}_i - {\vec k}_j)\,
({\vec \tau}_i \times {\vec \tau}_j)_z \nonumber \\
&{}& \frac{(\Lambda_{\pi}^2-m_{\pi}^2)^2}{({\vec k}_i^2+\Lambda_{\pi}^2)
({\vec k}_j^2+\Lambda_{\pi}^2)}\, 
\left(1 + \frac{{\vec k}_i^2+m_{\pi}^2}{{\vec k}_j^2+\Lambda_{\pi}^2}+
\frac{{\vec k}_j^2+m_{\pi}^2}{{\vec k}_i^2+\Lambda_{\pi}^2}\right) \, ,
\label{eq:gammaqq}
\end{eqnarray}
where $\vec k_i, \vec k_j$ are the momenta delivered to quarks $i$ and $j$ with
mass $m_i, m_j$, respectively, and the momentum conservation reads 
$\vec q= \vec k_i + \vec k_j$. The parameters $m_{\pi}, g_{\pi}, 
\Lambda_{\pi}$ are the mass, the coupling constant and the cut-off of the 
pion-quark vertex parametrized as 
\begin{equation}
F(\vec q) = {{\Lambda^2-m^2}\over{\Lambda^2+\vec q^2}} \, .
\label{eq:monopole}
\end{equation}

Analogously, the vector piece gives the well known isovector 
$\rho$-pair $(\rho q \overline{q})$ and $\rho$-in-flight 
$(\gamma \rho \rho)$ currents~\cite{ts99}
\begin{eqnarray}
\vec J_{\rho q\bar q}(\vec k_i, \vec k_j) &=& {\mathrm{i}} \,
\frac{(g_{\rho}^V+g_{\rho}^T)^2}{4m_i m_j}\left[
\frac{{\vec\sigma}_i \times ({\vec\sigma}_j \times {\vec k}_j)}
{({\vec k}_j^2+m_{\rho}^2)}\,\left(
\frac{\Lambda_{\rho}^2-m_{\rho}^2}{{\vec k}_j^2+\Lambda_{\rho}^2}\right)^2-
\, ( i \leftrightarrow j )\right] \nonumber \\
 &{}& \qquad\,({\vec \tau}_i \times {\vec \tau}_j)_z 
\label{eq:rhoqq}
\end{eqnarray}
\begin{eqnarray}
\vec J_{\gamma \rho\rho}(\vec k_i, \vec k_j) &=& {\mathrm{i}} \,\left[
(g_{\rho}^V)^2+\frac{(g_{\rho}^V+g_{\rho}^T)^2}{4m_i m_j}\,
({\vec\sigma}_i\times {\vec k}_i)\cdot({\vec\sigma}_j\times {\vec k}_j)\right]
\frac{({\vec k}_i - {\vec k}_j)}
{({\vec k}_i^2+m_{\rho}^2)({\vec k}_j^2+m_{\rho}^2)}\, \nonumber \\
&{}& \frac{(\Lambda_{\rho}^2-m_{\rho}^2)^2}{({\vec k}_i^2+\Lambda_{\rho}^2)
({\vec k}_j^2+\Lambda_{\rho}^2)}\, 
\left(1 + \frac{{\vec k}_i^2+m_{\rho}^2}{{\vec k}_j^2+\Lambda_{\rho}^2}+
\frac{{\vec k}_j^2+m_{\rho}^2}{{\vec k}_i^2+\Lambda_{\rho}^2}\right)\,
({\vec \tau}_i \times {\vec \tau}_j)_z  \, .
\label{eq:gammarhorho}
\end{eqnarray}
The parameters $m_{\rho}, \Lambda_{\rho}, g_{\rho}^V, g_{\rho}^T$ are the 
$\rho$ mass, cut-off, vector and tensor coupling constants of the $\rho$-quark 
vertex, respectively. All the parameter values have been kept the same as the 
ones used in Refs.~\cite{exGBECQM,exGBECQM1} for reproducing the baryon 
spectrum.

We have also explored the role of ``model-dependent'' two-body currents,
namely of operators which are not constrained by the continuity equation
(\ref{eq:cont}) because of their transverse nature. In particular, we have
considered the well known isoscalar $\rho \pi \gamma$ current
\begin{eqnarray}
\vec J_{\rho\pi\gamma}(\vec k_i, \vec k_j) &=& {\mathrm{i}} {{g_{\pi}}
\over 2m} {{g_{\rho}^V}\over m_{\rho}} g_{\rho\pi\gamma} \left[ {{\sigma_i
\cdot \vec k_i}\over {(\vec k_i^2+m_{\pi}^2)(\vec k_j^2+m_{\rho}^2)}} \, 
{{\Lambda_{\rho}^2-m_{\rho}^2}\over {\vec k_j^2+\Lambda_{\rho}^2}} \, 
{{\Lambda_{\pi}^2-m_{\pi}^2}\over {\vec k_i^2+\Lambda_{\pi}^2}} - \, ( i
\leftrightarrow j ) \right] \nonumber \\
&{}&\quad  \vec k_i \times \vec k_j \, \,  \vec \tau_i \cdot \vec \tau_j \,
, \label{eq:rhopigamma}
\end{eqnarray}
where $g_{\rho\pi\gamma}=0.578\pm 0.028$, in accordance with the Vector Meson
Dominance hypothesis (VMD) for the $\rho\rightarrow \pi\gamma$ decay 
width~\cite{Towner}.

The magnetic moment is the global sum of the contributions corresponding to
each previous component of the current operator:
\begin{equation}
\mu_N = \mu_N^{\left[ 1\right]} + \mu_N^{\pi q \overline{q}} +
\mu_N^{\gamma\pi\pi} + \mu_N^{\rho q \overline{q}} + \mu_N^{\gamma\rho\rho} +
\mu_N^{\rho\pi\gamma} \, .
\label{eq:magnmom}
\end{equation}
In Table (\ref{tab1}) the different results are shown. The one-body
contribution is the leading one, as expected, but the proper treatment of the cm
and intrinsic Hamiltonians represents a substantial improvement with respect to
Refs.~\cite{prc,ts99} and leads to a very good reproduction of the 
experimental values. The isovector two-body contributions, constrained by the 
continuity equation, show large cancellations but are globally important and 
act in opposite and correct ways according to the nucleon isospin. Finally, the 
isoscalar $\rho\pi\gamma$ contribution, though small, adds with the same sign 
both to proton and neutron magnetic moments, thus reducing the deviation of the 
theoretical values from the observed ones. The net theoretical result is in good 
agreement with the experiment, specifically with an error of about $1.5\%$ 
for the proton and of about $2\%$ for the neutron.

\begin{table}[h]
\caption{Contributions to the magnetic moments of the proton and neutron from 
different currents.\label{tab1}}
\vspace{0.2cm}
\begin{center}
\footnotesize
\begin{tabular}{|c|c|c|c|c|c|c|c|c|}
\hline
N & \qquad \qquad \qquad $\mu_N^{\left[ 1\right]}$ & $\mu_N^{\pi q 
\overline{q}}$ & $\mu_N^{\gamma \pi\pi}$ & $\mu_N^{\rho q\bar q}$ &
 $\mu_N^{\gamma \rho\rho}$ & $\mu_N^{\rho\pi\gamma}$ & $\mu_N$ & exp 
 \\ \hline
 p&\begin{tabular}{c c}
   u & 1.773 \\
   d & 0.215 \end{tabular} \Bigg \}
  1.988 & -0.126 & 0.737 & -0.109 & 0.202 & 0.048 & 2.740 & 2.793 \\ \hline
 n&\begin{tabular}{c c}
   u & -0.430 \\
   d & -0.886 \end{tabular} \Bigg \}
  -1.316 & 0.126 & -0.737 & 0.109 & -0.202 & 0.048 & -1.972 & -1.913 \\ 
\hline
\end{tabular}
\end{center}
\end{table}

\vspace{.5 cm}

We are grateful to Willi Plessas for useful discussions. This work was partly 
performed under the contract ERB FMRX-CT-96-0008 within the frame of the 
Training and Mobility of Researchers Programme of the Commission of the 
European Union.

%%%%%%%%%%%%%%%%%%%%%%%%%%%%% bibliography %%%%%%%%%%%%%%%%%%%%%%%%%%%

\end{document}